\documentclass[runningheads]{llncs}
\usepackage{url}
\usepackage{graphicx}
\usepackage{amsmath,amssymb,latexsym,bm,xspace}
\usepackage{multirow}
\usepackage{array}
\usepackage{makecell}
\newcommand{\vadl}{$\ell_{\text{\sf vol-dice}}$\xspace}
\newcommand{\dice}{$\ell_{\text{\sf dice}}$\xspace}
\newcommand{\rbr}[1]{\left(#1\right)}
\newcommand{\vol}[1]{{\sf volume}\left(#1\right)}
\newcommand{\bp}{{\bm{p}}}
\newcommand{\bg}{{\bm{g}}}

\newcommand{\repeatthanks}{\textsuperscript{\thefootnote}}

\begin{document}
\title{Multimodal Volume-Aware Detection and Segmentation for Brain Metastases Radiosurgery}

\titlerunning{Multimodal Volume-Aware Brain Mets Segmentation}
%
%
\author{Szu-Yeu, Hu\inst{1}\thanks{work done while at Vysioneer}, Wei-Hung Weng\inst{2}\repeatthanks, Shao-Lun Lu\inst{3}, Yueh-Hung Cheng\inst{4}, Furen Xiao\inst{5}, Feng-Ming Hsu\inst{3}, Jen-Tang Lu\inst{4}}
\authorrunning{S-Y. Hu et al.}
%

\institute{Massachusetts General Hospital, Boston, MA, USA \and Massachusetts Institute of Technology, Cambridge, MA, USA \and Department of Oncology, National Taiwan University Hospital, Taipei, Taiwan, \and Vysioneer Inc., Cambridge, MA, USA \and Department of Surgery, National Taiwan University Hospital, Taipei, Taiwan\\
\email{jt@vysioneer.com}}

\maketitle  
\begin{abstract}
Stereotactic radiosurgery (SRS), which delivers high doses of irradiation in a single or few shots to small targets, has been a standard of care for brain metastases. While very effective, SRS currently requires manually intensive delineation of tumors. In this work, we present a deep learning approach for automated detection and segmentation of brain metastases using multimodal imaging and ensemble neural networks. In order to address small and multiple brain metastases, we further propose a volume-aware Dice loss which optimizes model performance using the information of lesion size. This work surpasses current benchmark levels and demonstrates a reliable AI-assisted system for SRS treatment planning for multiple brain metastases.


\end{abstract}

\section{Introduction}
Brain metastases (BMs) are the most common intracranial tumors in adults (10 times more common than primary brain tumors) and occur in around 20\% of all patients with cancer \cite{lin2015treatment}. 
The treatment options for brain metastases include craniotomy, chemotherapy, whole brain radiation therapy, and stereotactic radiosurgery (SRS). Among all the options, SRS has been playing a critical role in the treatment of brain metastases as recent studies have shown that SRS leads to better treatment outcomes \cite{tsao2012radiotherapeutic}. By delivering high doses of irradiation in a single or few shots to small targets, SRS effectively destroys tumors without damaging surrounding tissues and has been proved to be beneficial in the local tumor control and post-operative neurocognitive function \cite{hartgerink2018stereotactic}.

As SRS requires precise delineation of tumor margins, target segmentation (contouring) for BMs is performed manually by the radiation oncologist or neurosurgeon on magnetic resonance images (MRI) and computed tomography (CT) images of the brain. However, such manually contouring process can be very time-consuming and suffer from large inter and intra-reader variability \cite{vinod2016uncertainties}. Driven by the ever-increasing capability of deep learning, automated segmentation of BMs using neural networks has been recently proposed \cite{charron2018automatic,liu2017deep}. Previous works on computer-aided segmentation of BMs used only MRI as an imaging input. While MRI provides superior ability to characterize neural tissue and the brain structure, MRI is prone to have the problems of spatial distortion and motion artifacts, which can lead to inaccuracy in SRS. CT, on the other hand, is lack of soft tissue contrast but provides a direct measurement of electron densities for radiation dose calculations and has excellent spatial fidelity. Consequently, co-registration between MRI and CT modalities is recommended for precise stereotactic applications \cite{pereira2014role}.


Many previous works on automated segmentation of brain tumors focused on multiforme glioblastoma and glioma, such as BraTS dataset \cite{menze2015multimodal}, and aims to optimize Dice similarity coefficient (DSC) \cite{milletari2016v}. Segmentation of brain metastases is more challenging as metastatic lesions can be very small ($<$ 1000 $mm^3$) and a large brain metastasis can coexist with multiple small lesions. Conventional DSC is thus not an ideal metric to evaluate brain metastases segmentation because it would be dominant by the large lesion but ignore small metastases. Unfortunately, small BMs are much crucial to SRS since they are more likely to be missed by clinicians.


In this paper, we aim to utilize deep neural networks for automated detection and segmentation of brain metastases. Specifically, we present a deep learning-based system for brain metastases detection and segmentation using multimodal imaging (MRI+CT) and ensemble neural networks, which produces a more reliable result than that would be achieved by a single image modality and/or a single neural network. To address the challenge of small BMs, we further propose a volume-aware Dice loss (\vadl), which leverages the information of lesion size to optimize overall segmentation.

\section{Methods}

\subsection{Volume-Aware Dice Loss}
The Dice loss (\dice), which aims to optimize DSC, has been widely used as a loss function in medical image segmentation task~\cite{milletari2016v} and can be expressed as:
\vspace{-0.5em}
\begin{equation}
  \text{\dice}\rbr{\bg, \bp} =  - \frac{2 \bg^\top\bp +
  \epsilon}{\bp^\top\bp + \bg^\top\bg + \epsilon},
\end{equation}
where $\bp \in [0,1]^N$ and $\bg \in \{0, 1\}^N$ are the predicted probability
vector and the ground truth binary vector for $N$ voxels, respectively. $\epsilon$ is a smoothing constant to avoid the denominator being zero.

For binary segmentation problems with similar sizes of lesions, using \dice as the objective function is a fair option as it is normalized by the number of foreground pixels. However, \dice is not an ideal choice when multiple targets are present with the same label but with different sizes, since the loss will be dominant by the larger targets. To address the issue, we proposed a Volume-Aware Dice Loss (\vadl) that \textit{optimizes the overall segmentation using the information of lesion size}. \vadl can be formulated as:
\begin{equation}
  \text{\vadl}\rbr{\bg, \bp\mid W} =
    - \frac{C \bg^\top W \bp + \epsilon}
    {\bp ^\top \bp + \bg^\top W \bg + \epsilon},
\end{equation}
where $W \in \mathbb{R}^{N\times N}$ is a diagonal matrix that $W_{ii}$ is a weight related to the volume of the tumor containing the $i$-th voxel in the ground truth, denoted by $\vol{i}$. $\textstyle C:= 1 + {\bg^\top\bg}/{\bg^\top W \bg}$ is a normalization constant such that the maximum of \vadl is one. In this paper, we consider the form:
\[
  W_{ii} = \begin{cases}
    \rbr{\textstyle\frac{\lambda}{\vol{i}}}^{1/2} \ & \vol{i} \neq 0, \\
    0 & \text{otherwise}
  \end{cases},
\]
where $\lambda$ is a reweighting hyper-parameter and can be defined as follows.
\begin{enumerate}
    \item \textbf{Constant Reweight} (CR): $\lambda$ is a constant ($\lambda_c$), such that the weights are simply proportional to the inverse of the square root of tumor volume.
    \item \textbf{Batch Reweight} (BR): $\lambda$ is the largest tumor volume in each batch ($\lambda_l$); e.g. if the largest tumor in one batch has 1500 voxels, another small tumor with 60 voxels will have a weight of $\sqrt{1500/60} = 5$.
\end{enumerate}

To demonstrate the effects of \vadl, assume a brain image volume with three ground-truth tumors, each with 1800, 450 and 200 voxels. Using the batch reweight \vadl, each tumor will have weights 1, 2 and 3, respectively. Suppose the model perfectly predicts the two larger tumors but fails to detect the smallest one. Under such a scenario, \dice is calculated as $-\frac{2\times (1800 + 450)}{(1800 + 450 + 200) + (1800 + 450)} = -0.957$ and \vadl is $-\frac{C \times (1800 + 450 \times 2)}{(1800 + 450 \times 2 + 200 \times 3) + (1800 + 450)} = -0.847$, where $C = 1 + \frac{2450}{3300}$. It illustrates that \vadl is more sensitive to the small structures. 


\subsection{Deep Learning Framework}
\begin{figure}[t!]
\centering
\includegraphics[width=1.0\linewidth]{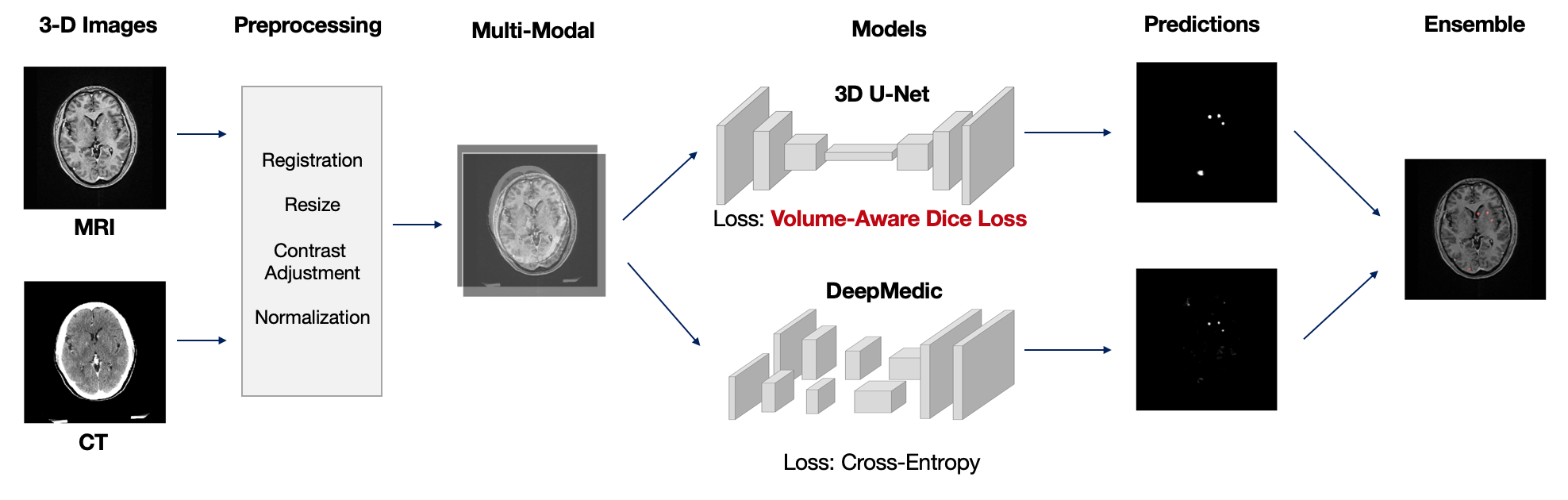}
\hfill
\vspace{-2.5em}
\caption{Proposed deep learning framework with multimodal imaging and ensemble networks.}
\label{fig:fig1}
\vspace{-10pt}
\end{figure}

While the current benchmark for brain metastases segmentation employs MRI imaging only \cite{charron2018automatic,liu2017deep}, CT imaging is an essential reference for clinical treatment planning due to its spatial accuracy. We thus proposed a deep learning framework using multimodal imaging (MRI+CT) and ensemble neural networks for brain metastases detection and segmentation. The framework is shown in Figure \ref{fig:fig1}.  For the ensemble model, we explored two different architectures---3D U-Net and DeepMedic.

\subsubsection{3D U-Net}
3D U-Net \cite{cciccek20163d} is an extension of the original U-Net by replacing all the 2D operations with 3D counterparts. We added one block in addition to the original implementation (number of feature maps: 32, 64, 128, 256, 512 with convolution kernel size 3$\times$3$\times$3 and max-pooling size 2$\times$2$\times$1). We took a full size of the axial-view images and randomly sampled 8 consecutive slices on the vertical axis, resulting in input images size of 512$\times$512$\times$8$\times$2 (height $\times$ width $\times$ number of slices $\times$ number of imaging modality). We set a limit to ensure that in each epoch, at least 70\% of the samples should contain tumor labels. All the 3D U-Net models were trained using a \textsf{rmsprop} optimizer, with learning rate $10^{-3}$, batch size of 1, over 300 epochs on an NVIDIA V100 GPU. The best weights on the validation set was used to evaluate the final results on the test set.
\vspace{-1em}
\subsubsection{DeepMedic}
DeepMedic~\cite{kamnitsas2017efficient} is originally designed for brain tumor segmentation on multi-channel MRI and also had been applied to BMs~\cite{charron2018automatic,liu2017deep}. It consists of multiple parallel pathways---one branch takes small patches from full resolution images as input and the others utilizes subsampled-version of the images. Different from the original paper, the DeepMedic architecture we used contained 3 parallel convolutional pathways, one of which was with normal image resolution and the other two of which were with low resolution using down-sampling factors of 3 and 5. There were 11 layers in the network, the first 8 of which were convolutional layers (number of feature maps: 30, 30, 40, 40, 40, 40, 50, 50 with 3$\times$3$\times$3 kernels) and the last 3 of which were fully connected layers (with 250 feature maps per layer). The network was trained using a \textsf{rmsprop} optimizer, with learning rate $10^{-3}$, minimizing the cross-entropy loss over 35 epochs.
\vspace{-1em}
\subsubsection{Ensemble Model}
3D U-Net and DeepMedic were trained separately with different hyperparameters and different objective functions to maximize the capability of the ensemble model. While U-Net utilized the full field of view for each image slice and addressed overall tumor segmentation, DeepMedic leveraged image segments during model training and focused on small metastases. Furthermore, U-Net was trained to optimize DSC and DeepMedic was set to minimize cross-entropy loss. At testing time, each model individually generated probability maps of brain metastases. An ensemble confidence map was then created by calculating the average of the predictions of both models.



\section{Experiments and Results}

\subsection*{Data}
The primary cohort for model training and testing consists of 305 patients with 864 brain metastases (median volume 760 $mm^3$, range 3-110,000 $mm^3$) treated by CyberKnife G4 system in a single medical center. Each case contains volume masks of brain tumors delineated by an attending radiation oncologist or neurosurgeon on associated CT and T1-weighted MRI scan with contrast. The dataset was split into training (80\%), validation (10\%),  and test set (10\%) randomly. To evaluate the model robustness and generalizability, we collected an additional batch of test set from the same institution, containing 36 patients with 96 metastases (median volume 829 $mm^3$). The results presented in this paper are the average of the two test sets.

For each case, after rigid image registration between CT and MRI image volumes, each slice was resized into 512$\times$512 pixels with the resolution of 0.6mm $\times$ 0.6mm, while slice thickness was resampled to 2mm. Brain window and adaptive histogram equalization were applied to the CT and MRI images slice-by-slice, respectively. All the image volumes were then standardized with zero-mean and unit-variance normalization.

\vspace{-1em}
\subsection{Volume-Aware Dice Loss}

\subsubsection{\vadl with different reweight strategies}
We evaluated the efficacy of \vadl on the 3D U-Net. A standard 3D U-Net using the multimodal learning (MRI+CT) and conventional \dice was trained as the baseline model. Then we compared the relative change of DSC, precision and recall using different settings of \vadl. In the combination of two test sets, the tumors has median 1322 voxels (949 $mm^3$) and mean 4721 voxels (3389 $mm^3$). We tested the $\lambda_c$ of 500, 1000, 2500, 5000 for the CR strategy.

Our baseline model achieves a DSC of 0.669, precision 0.689 and recall 0.700. Table \ref{tab:table3} lists the results of applying \vadl relative to the baseline. Overall, using the \vadl with BR ($\lambda_l$) yields the best performance, improving 8.57\% of DSC and 24.14\% of recall compared to the baseline. The performance of \vadl with CR largely depends on the constant value; the higher the constant, the better the recalls. Such an observation is consistent with our expectation that the \vadl focuses more on easily neglected small structures and has a higher sensitivity. However, in our dataset, the tumor sizes are highly diverse, therefore making it challenging to determine a value that can generalize to all the tumors. On the other hand, the BR approach has more flexibility using a dynamic weighting strategy, which provides a balance between precision and recall.

\begin{table}[t!]
\centering
\caption{Performance of different configurations with \vadl. All the models were trained on 3D U-Net with CT and MRI.}
\label{tab:table3}
\begin{tabular}{c|c|r|r|r}
\hline
 \multicolumn{2}{c|}{\vadl configuration} &\multicolumn{3}{c}{Metric Change over \dice loss (\%)} \\
\hline
Reweight method & $\lambda$ & DSC & Precision & Recall \\ 
\hline
\hline
Batch-Reweight  & - & \textbf{+8.57\%} & +2.62\% & +24.14\%\\
Const-Reweight & $\lambda_c = 500$ & -52.11\% & +2.78\% & -70.08\%\\
Const-Reweight & $\lambda_c = 1000$ & -7.67\% & \textbf{+13.15\%} & -18.83\% \\
Const-Reweight & $\lambda_c = 2500$ &+2.22\% & -5.55\% & +23.04\% \\
Const-Reweight & $\lambda_c = 5000$ & -8.60\% & -31.68\% & \textbf{+25.46\%} \\
\hline
\end{tabular}

\end{table}

\begin{table}[!t]
\centering
\caption{Pixel-wise and metastasis-wise recall}
\label{tab:table5}
\begin{tabular}{c|c|c|c|c|c}
\hline
Tumor & Numbers & Median Size & Loss & Pixel-wise recall & Metastasis-wise recall \\ \hline
\multirow{2}{*}{\makecell{Small \\ ($\leq$ 1500 mm$^3$)}} & \multirow{2}{*}{113} & \multirow{2}{*}{368} & \dice & 0.466 & 0.619\\ 
& & & \vadl & 0.633 & 0.672\\ \hline
\multirow{2}{*}{\makecell{Large \\ ($>$ 1500 mm$^3$)}} & \multirow{2}{*}{78} & \multirow{2}{*}{4114} & \dice & 0.826 & 0.987 \\ 
& & & \vadl & 0.828 & 0.974 \\ \hline

\end{tabular}
\vspace{-1em}
\end{table}

\subsubsection{\vadl on small tumors}
To evaluate the effectiveness of \vadl on different sizes of lesions, we further divided the lesions into large and small tumor groups at a cut-off point of 1500 $mm^3$. We calculated (1) pixel-wise recall, and (2) tumor-wise recall, where a positive tumor prediction is defined as detected if there is at least one pixel being predicted; noted that we did not measure the DSC and the precision because the false positive pixels can't be categorized into either small or large tumors easily. The results are shown in Table \ref{tab:table5}. \vadl shows a more significant improvement in the small tumor groups, increasing the recall from 0.466 to 0.633 and detection rate from 0.619 to 0.672; while in the large tumor groups, the performances of the two settings are almost identical. The results indicate that the \vadl effectively improve the recall in the small tumors. 

\subsection{Deep Learning Framework}

In our final proposed deep learning framework, we used (1) multimodal learning adopting MRI+CT, (2) ensemble learning considering 3D U-Net and DeepMedic, and (3) optimization using \vadl. 

\begin{figure}[t!]
\begin{center}
\includegraphics[width=0.8\linewidth]{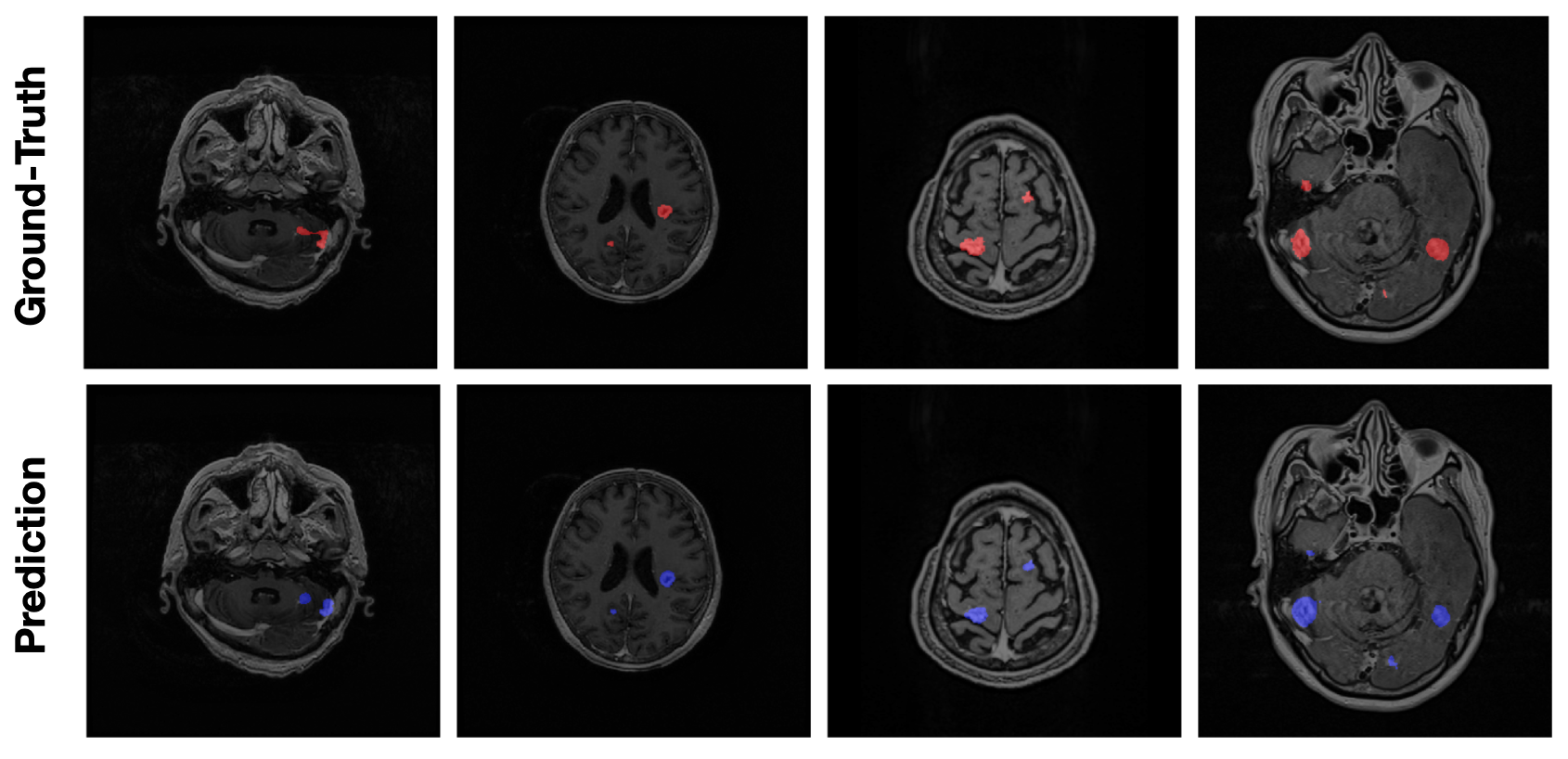}
\vspace{-1.5em}
\end{center}
\hfill
\vspace{-1.5em}
\caption{Examples of the prediction results overlaying with MRI.}
\vspace{-1em}
\label{fig:fig4}
\end{figure}

As shown in Table \ref{tab:table4}, 3D U-Net and DeepMedic obtain a DSC of 0.669 and 0.625 respectively. DeepMedic utilizes a patch-based training method, which makes the network focusing on smaller regions and contributing to a higher recall; on the other hand, 3D U-Net takes full resolution as inputs. The advantage of seeing the complete brain structure leads to higher precision. The ensemble of the two models improves the DSC to 0.719. Further applying the \vadl (with BR) on 3D U-Net, we achieve the best DSC of 0.740 and recall of 0.803. The results indicate that our deep learning approach effectively increases the DSC, precision, and recall. The performance surpasses the current benchmark methods. Examples of prediction results are shown in Figure~\ref{fig:fig4}.

\begin{table}[th!]
\vspace{-1em}
\centering
\caption{Model performance of different configurations of loss functions, image modalities, and neural network models. The values are represented as median (std).}
\label{tab:table4}
\vspace{-1em}

\begin{tabular}{c|c|c|c|c}
\hline
Model & \vadl & DSC & Precision & Recall \\ \hline
 3D U-Net & & 0.669 (0.006) & 0.689 (0.001) & 0.700 (0.015) \\
DeepMedic & & 0.625 (0.013)& 0.631 (0.004)& 0.734 (0.035) \\
3D U-Net + DeepMedic & &0.719 (0.004) & \textbf{0.788} (0.002) & 0.713 (0.023) \\
3D U-Net + DeepMedic & \checkmark &\textbf{0.740} (0.022) & 0.779 (0.010)& \textbf{0.803} (0.001) \\ \hline

\end{tabular}
\vspace{-0.2em}
\end{table}

\vspace{-1em}
\subsection{Limitation}
\begin{figure}[t!]
\begin{center}
\includegraphics[width=0.8\linewidth]{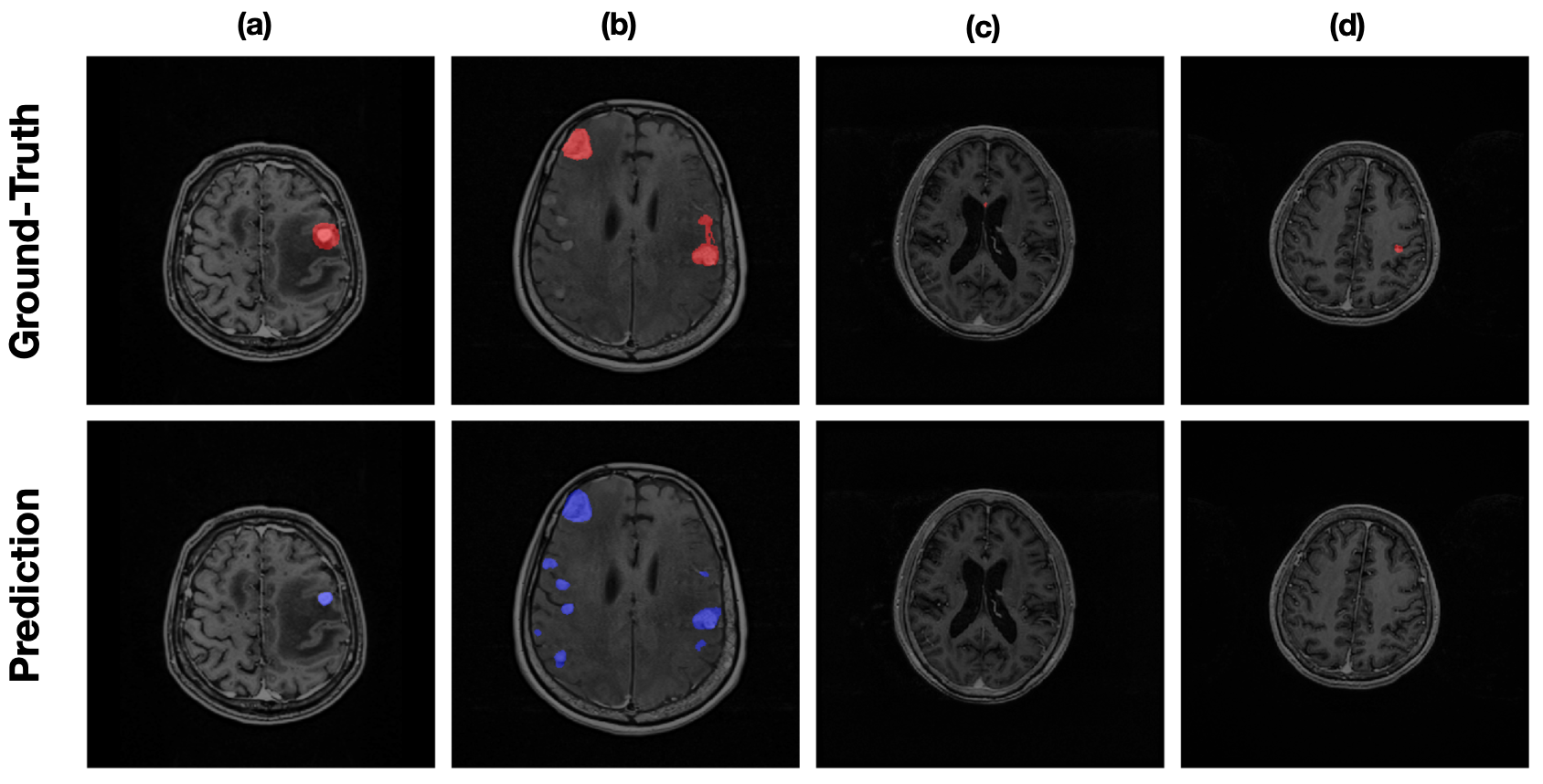}
    \vspace{-1.5em}
\end{center}
\hfill
\vspace{-1.5em}
\caption{Examples of failed cases overlaying with MRI.}
\vspace{-2em}
\label{fig:fig5}
\end{figure}

As the annotations were carried out by the neurosurgeon or radiation oncologist during SRS treatment planning, the ground truth labels represent the area for the treatment rather than the actual tumor extent, which leads to imperfect annotations for tumor segmentation. Based on the clinician's experience and the patient's disease status, these annotations can be delineated more aggressively or conservatively. Figure \ref{fig:fig5}(a) illustrates an example of an aggressive treatment planning, which shows a broader area than the lesion. Also,the clinician would ignore previously treated tumors (Figure \ref{fig:fig5}(b)). Last, some difficult cases, such as Figure \ref{fig:fig5}(c) and (d), are highly subjective and should be determined through clinical manifestations or the series change of MRI. The cases mentioned above can underestimate our model performance and lead to a higher variance.

%
\vspace{-1em}

\section{Conclusion}
In this paper, we have achieved high performance for automated detection and segmentation of brain metastases, utilizing multimodal imaging (MRI+CT) as inputs and ensemble neural networks. We have also addressed the challenge of lesion size variance in multiple metastases by introducing a volume-aware Dice loss, which leverages the information of lesion size and significantly enhances the overall segmentation and sensitivity of small lesions, which are critical in the current SRS contouring workflow. It is expected that the proposed solution will facilitate tumor contouring and treatment planning of stereotactic radiosurgery.

\bibliographystyle{splncs04}
\bibliography{ref}
\end{document}